\documentclass[final,1p,times]{elsarticle} 

\usepackage{graphicx}
\usepackage{amssymb} 
\usepackage{amsthm} 
\usepackage{lineno}
\usepackage{xspace}
\newcommand{\RAA}{\ensuremath{R_{\rm{AA}}}\xspace}
\newcommand{\pp}{\ensuremath{\rm pp}\xspace}
\newcommand{\pbpb}{\ensuremath{\rm Pb-Pb}\xspace}

\newcommand{\pt}{\ensuremath{p_{\rm{T}}}\xspace}
\newcommand{\mdedx}{\ensuremath{\langle {\rm d}E/{\rm d}x \rangle}\xspace}
\newcommand{\dedx}{\ensuremath{{\rm d}E/{\rm d}x}\xspace}
\newcommand{\dpi}{\ensuremath{\Delta_{\pi}}\xspace}
\newcommand{\gev}{\ensuremath{{\rm GeV}/c}\xspace}

\journal{Nuclear Physics A} 

\begin{document}

\begin{frontmatter} 

\title{Production of pions, kaons and protons at high $p_{\rm T}$ in $\sqrt{s_{\rm NN}}=$ 2.76 TeV Pb-Pb collisions}

\author{Antonio Ortiz Velasquez (for the ALICE\fnref{col1} Collaboration)}
\fntext[col1] {A list of members of the ALICE Collaboration and acknowledgements can be found at the end of this issue.}
\address{Division of Particle Physics, Lund University, Sweden}


\begin{abstract} 
In this work we report on the production of charged pions, kaons, and (anti) protons in \pp and \pbpb collisions at $\sqrt{s_{NN}}$ = 2.76 TeV for $3 < p_{\rm T} < 20$ \gev. The evolution of the nuclear modification factor, $R_{AA}$,  with collision centrality and transverse momentum is discussed. 
\end{abstract} 

\end{frontmatter} 


\section{Introduction}
\label{ch:1}

The particle identification (PID) is an important tool to study the hot and dense matter created in relativistic heavy ion collisions. At intermediate \pt (2-8 \gev)  hadrons can be produced not only by fragmentation but also by quark recombination. It allows us to learn how the medium affects the particle composition~\cite{ref:qgp}. In Section 2 we describe the analysis to measure the production of $\pi/{\rm K/p}$ as a function of \pt (3-20 \gev). The discussion of the results and conclusions are presented in Sections 3 and 4, respectively.

\section{Identification of charged hadrons at high \pt}
\label{ch:2}
The measurement of the yields of charged pions, kaons and (anti) protons in the range $3<p_{\rm T}<20$ \gev is statistically performed exploiting the peculiarity of the mean energy loss, \mdedx, of particles traversing the the TPC gas~\cite{{ref:ppr1}}, in the relativistic rise regime ($\beta \gamma: 3.6-1000$) of the Bethe-Bloch (BB) curve.

The TPC response was calibrated using a clean sample of $\pi$ ($\rm p$) identified
through the weak decay topology of ${\rm K_{S}^{0}}$ ($\Lambda$). Primary pions and electrons were tagged using the Time of Flight (TOF) detector. After parametrization of the BB curve and of the \dedx resolution as a function of the calibrated \mdedx, the measured resolution of  minimum ionizing particles were $5.6\%$ and $7.5\%$ for low (\pp) and high (\pbpb 0-5\%) multiplicity environment, respectively.

The yields were extracted by fitting the quantity \dpi$={\rm d}E/{\rm d}x-\langle {\rm d}E/{\rm d}x \rangle_{\pi}$ to a sum of four Gaussian functions ($\pi/{\rm K/p}/e$) in each \pt interval. The means and sigmas were constrained using the aforementioned parametrization and the \pt spectra obtained via factorization, as described in~\cite{hp:peter}.

The main source of systematic uncertainty comes from the estimation of the parameters of the fitted functions.
A range of variation for those parameters was evaluated running the analysis without exploiting the information from topological and TOF PID. The parameters were randomly varied in the estimated ranges, then the yields were extracted using the same algorithm as in data. At high \pt ($>10$ \gev) the systematic uncertainties were 5\%, $10$\% and $\sim 30-40$\% for $\pi$, $\rm K$ and $\rm p$, respectively.  Also, we added the systematic errors from the spectra of inclusive charged particles~\cite{ref:chRAA}.

\section{Results}
\label{ch:3}

\begin{figure}[htbp]
  \begin{center}
    \includegraphics[keepaspectratio, width=0.95\columnwidth]{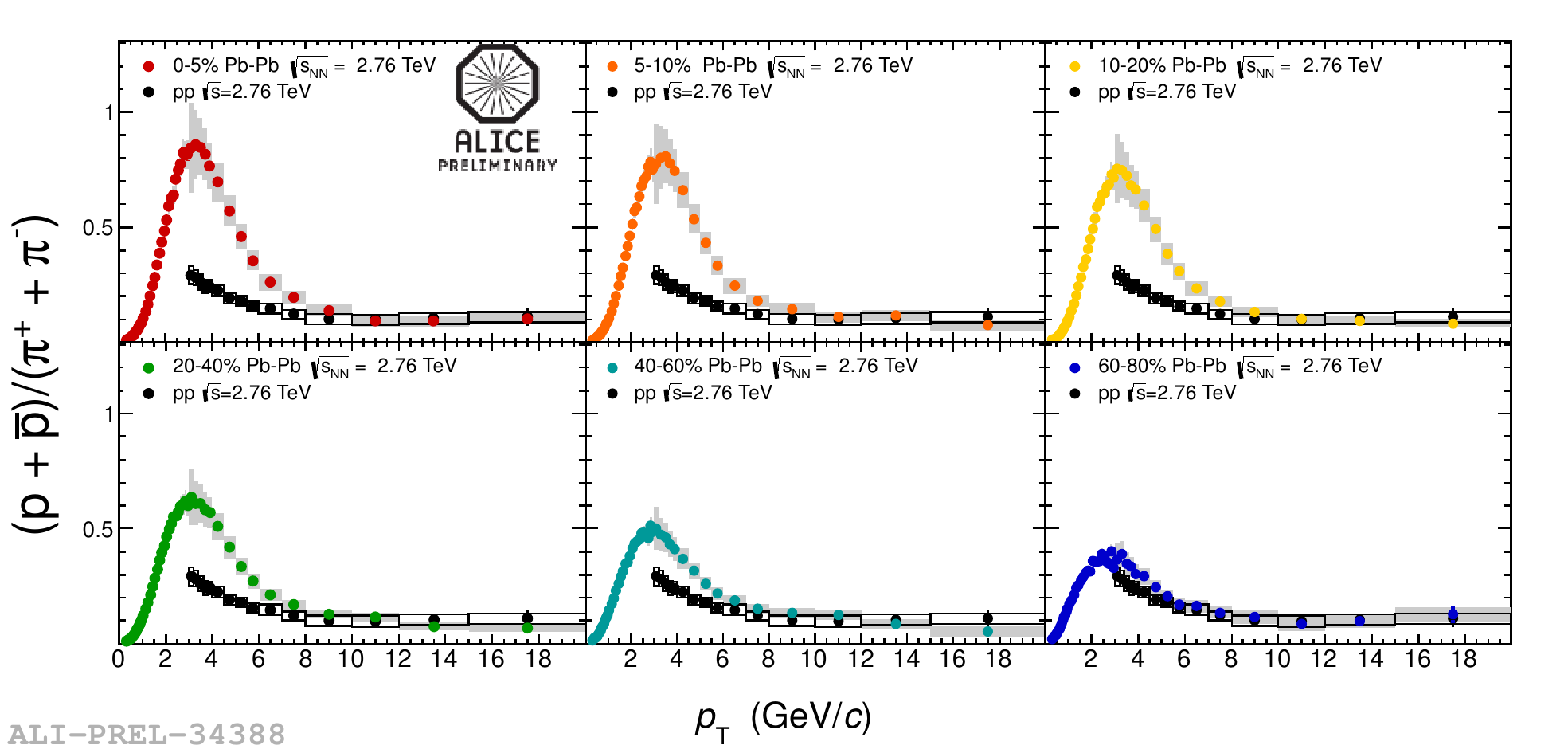}
  \end{center}
  \caption{Centrality dependence of the $({\rm p+ \bar{p}}) / (\pi^{+}+\pi^{-})$ ratio a a function of \pt compared to \pp at 2.76 TeV. Statistical and systematic uncertainties are displayed as error bars and boxes, respectively. The low \pt results come from the combined PID of ITS, TPC and TOF~\cite{ref:pidPbPbQM12}.}
  \label{fig:pTopi}
\end{figure}

\begin{figure}[htbp]
  \begin{center}
    \includegraphics[keepaspectratio, width=0.5\columnwidth]{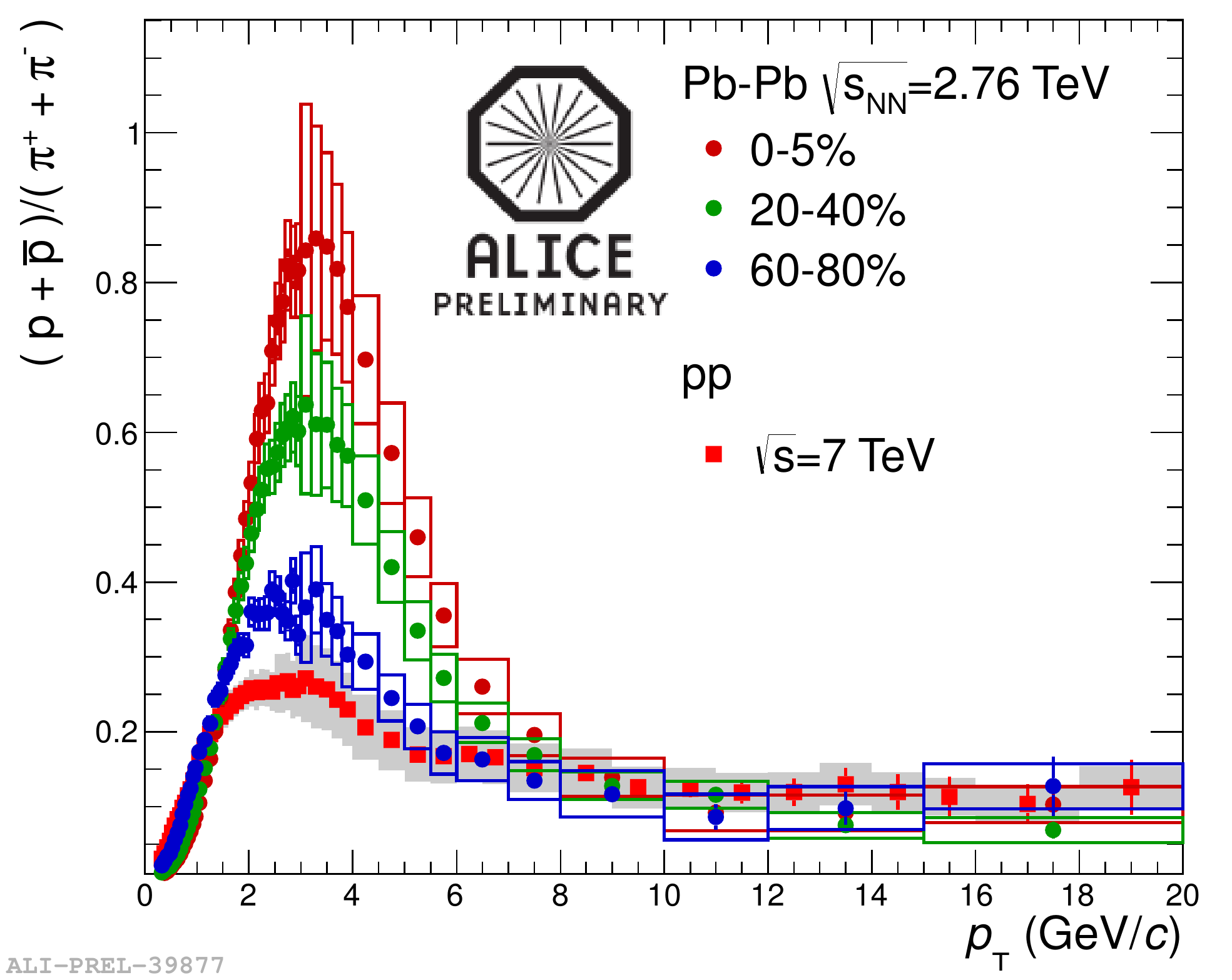}
  \end{center}
  \caption{Comparison of the $({\rm p+ \bar{p}}) / (\pi^{+}+\pi^{-})$ ratio as a function of \pt between \pbpb and \pp collisions at $\sqrt{s}_{\rm NN}=$ 2.76 TeV and 7 TeV, respectively. Statistical and systematic uncertainties are displayed as error bars and boxes, respectively. The low \pt results come from the combined PID of ITS, TPC and TOF~\cite{ref:pidPbPbQM12}. In addition the \pp result includes the information of the Cherenkov detector, HMPID.}
  \label{fig:pTopi7}
\end{figure}

Figure~\ref{fig:pTopi} shows $({\rm p+ \bar{p}}) / (\pi^{+}+\pi^{-})$ as a function of \pt for \pbpb and \pp collisions at $\sqrt{s_{\rm NN}}=$ 2.76 TeV. For completeness the lower \pt results are also displayed for \pbpb~\cite{ref:pidPbPb,ref:pidPbPbQM12}. At intermediate \pt the ratio is enhanced, this phenomenon was first observed at RHIC~\cite{ref:rhic1}. The maximum is located at 3 \gev and reaches  $\sim 0.8$ ($\sim 0.4$) for the most central (peripheral) events. For higher \pt ($>8$ \gev), the ratio is consistent with fragmentation in vacuum. Figure~\ref{fig:pTopi7} shows a comparison with the ratio in \pp collisions at $\sqrt{s}=$ 7 TeV. It is interesting that even the \pp data exhibit a maximum in the same  \pt region as observed in \pbpb collisions. 


\begin{figure}[htbp]
  \begin{center}
    \includegraphics[keepaspectratio, width=0.48\columnwidth]{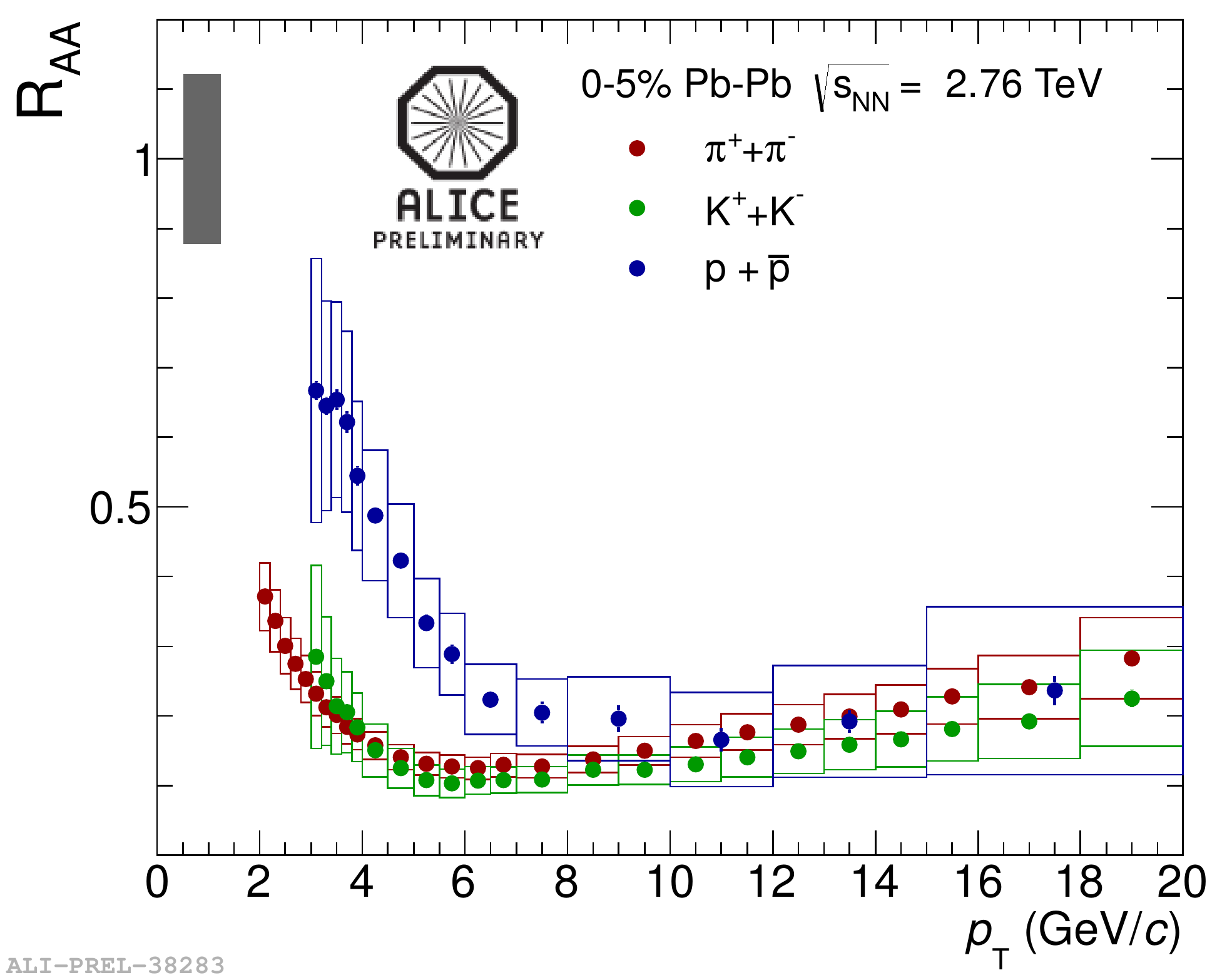}
    \includegraphics[keepaspectratio, width=0.48\columnwidth]{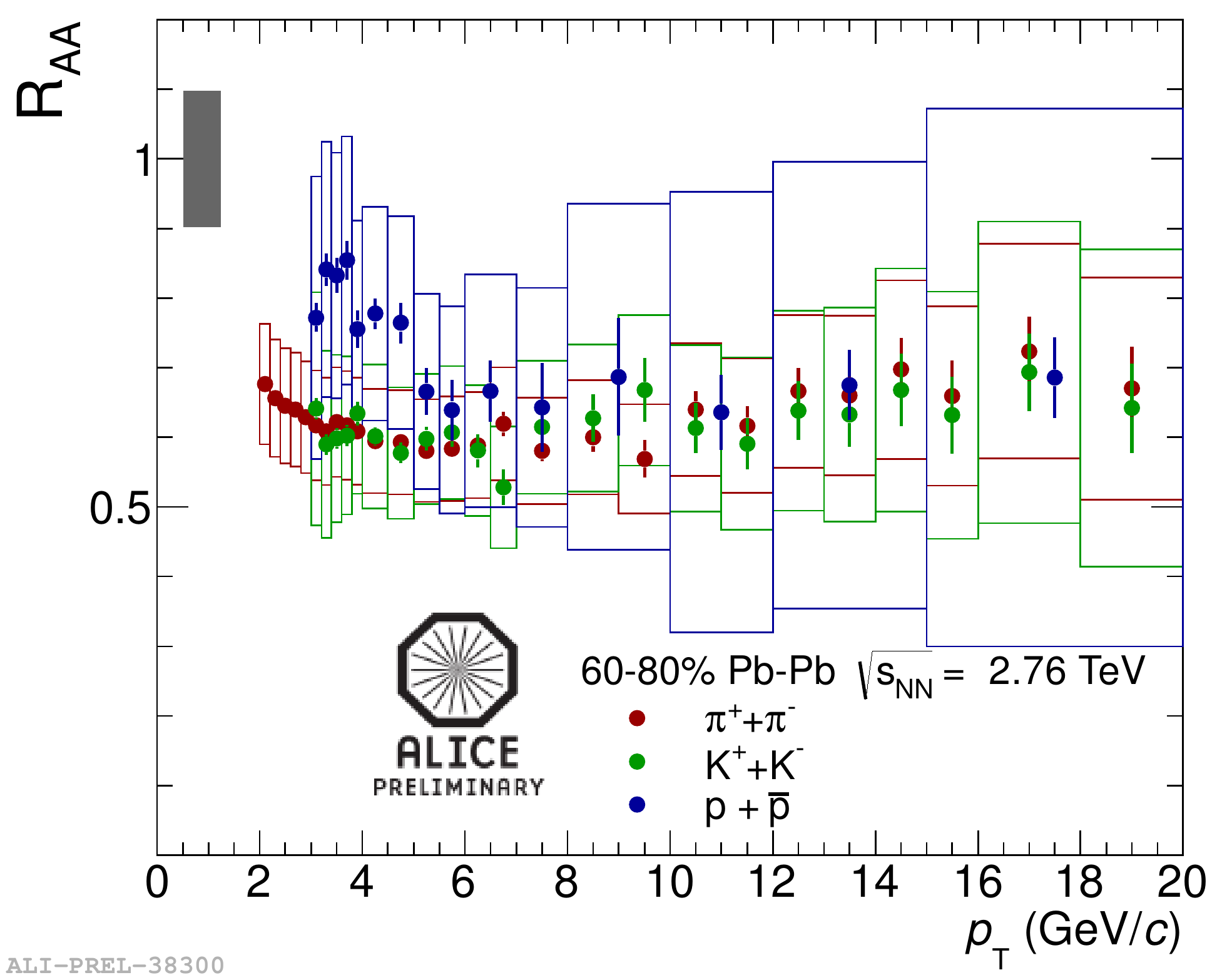}
  \end{center}
  \caption{\RAA as a function of \pt for $\pi/{\rm K/p}$ in 0-5\% (left) and 60-80\% (right) \pbpb collisions.  Statistical and systematic uncertainties are displayed as error bars and boxes, respectively.}
  \label{RAA:k}
\end{figure}

ALICE previously reported the  nuclear modification factor (\RAA) for $\pi$ and $\rm p+K$~\cite{hp:peter}. Now Figure~\ref{RAA:k} shows the \RAA as a function of \pt for pions, kaons and protons, the results are presented for 0-5\% and 60-80\%. For the most central events and  at intermediate \pt (2-8 \gev), \RAA$^{\pi}$ $<$ \RAA$^{\rm p}$ and  \RAA$^{\pi}$ $\sim$ \RAA$^{\rm K}$, no significant species dependence is seen at high \pt.

\section{Conclusions}
\label{ch:4}

At intermediate \pt, we observe an enhancement of the $({\rm p+\bar{p}})/(\pi^{+}+\pi^{-})$ ratio. Within systematics, $\pi/{\rm K/p}$ are equally suppressed at high \pt which suggests that the fragmentation is not modified by the medium.



\begin{thebibliography}{00} 
\bibitem{ref:qgp} K. Yagiet  {\it et al.,} ``Quark-Gluon Plasma'', Cambridge, (2005).
\bibitem{ref:ppr1}
  J.~Alme {\it et al.,} (ALICE Collaboration),
  Nucl.\ Instrum.\ Meth.\ A {\bf 622} (2010) 316.
\bibitem{hp:peter} Peter Christiansen, for the ALICE Collaboration. ``High \pt identified particle production in ALICE
'', arXiv:1208.5368.  
  
\bibitem{ref:chRAA} B. Abelev {\it et al.,} (ALICE Collaboration), ``Centrality dependence of charged particle production at large transverse momentum in \pbpb collisions at $\sqrt{s_{\rm NN}}=2.76$ TeV'', arXiv:1208.2711.

\bibitem{ref:pidPbPb}  B. Abelev {\it et al}.,  (ALICE Collaboration), ``Pion, Kaon, and Proton Production in Central \pbpb Collisions at $\sqrt{s_{\rm NN}} = 2.76$ TeV'', arXiv:1208.1974v1.
\bibitem{ref:pidPbPbQM12}  Leonardo Milano, for the ALICE Collaboration, ``Identified charged hadron production at the LHC with the ALICE experiment'', this proceedings.

\bibitem{ref:rhic1} S. Adler {\it et al}., Phys. Rev. Lett. 91, 172301 (2003).

\end{thebibliography}
\end{document}